\documentstyle[11pt,newpasp,twoside,epsf]{article}
\newcommand{\kms}{\ifmmode{\rm km\thinspace s^{-1}}\else km\thinspace s$^{-1}$\fi}

\begin{document}
\title{The Wisconsin H-Alpha Mapper Northern Sky Survey}
 \author{G.J. Madsen, L.M. Haffner, \& R.J. Reynolds}
\affil{Dept. of Astronomy, University of Wisconsin -- Madison, 475
  N. Charter Street, Madison, WI, 53706-1582}

\begin{abstract}
The Wisconsin H-Alpha Mapper (WHAM) has completed a one-degree
resolution, velocity-resolved northern sky survey of
{H$\alpha$}~emission from our Galaxy.
The unprecedented sensitivity of the instrument and accurate spectral
subtraction of atmospheric features allow us to detect Galactic
features as faint as 0.1 Rayleighs (EM $\approx 0.25$ cm$^{-6}$
pc). This survey
allows a direct comparison of the ionized and neutral components of
the ISM on a global scale for the first time. All-sky maps of
{H$\alpha$}~emission in select velocity bands highlight the rich  
kinematic structure of the Galaxy's ionized gas.
The full set of data from the WHAM survey is now available at
{\tt  http://www.astro.wisc.edu/wham/}.

One surprising result is that the high latitude sky of both the
ionized and neutral components display marked similarity in the
location and radial velocity of emitting regions, especially for many
of the previously 
identified intermediate velocity clouds. Although there is evidence
for spatial and velocity correlation, in many cases examined so far
there is little evidence for a quantitative correlation between the
column density of {H {\sc I}}~and the emission measure of {H {\sc II}}.

WHAM is also capable of studying the ISM through optical emission
lines other than H$\alpha$. Two directions toward the Perseus arm
have been studied in detail through several other optical emission
lines. The multiple velocity component structure toward
these directions provides a selection of ionized environments for
study and shows interesting variations in the ratios of these lines.
WHAM also has the ability to study select regions of the sky at high
spatial resolution (3$'$ to 5$'$) with high velocity resolution.  Such
observations will allow an even closer comparison between the
neutral and ionized components with the advent of high resolution
{H {\sc I}}~surveys.
\end{abstract}

\section{Introduction}

The WHAM instrument is a fully remotely operated facility with a
15 cm, dual-etalon Fabry-Perot spectrometer at the focal plane of a 0.6 m
telescope atop Kitt Peak in Arizona (Reynolds et al. 1998).
The WHAM spectrometer has a 1$^\circ$
diameter circular field of view on the sky, and a velocity
resolution of 12 km\thinspace s$^{-1}$~within a 200 km\thinspace
s$^{-1}$~wide spectral window that can 
be centered on any wavelength between 4800 \AA~and 7300 \AA. WHAM was
designed to detect very weak emission lines from ionized gas.

The Wisconsin H-Alpha Mapper Northern Sky Survey (WHAM-NSS) is the
first deep,velocity-resolved survey of interstellar
{H$\alpha$}~emission over the northern sky ($\delta \ge -30\deg$).
The survey consists of 37,565 individual observations taken over
a span of two years. Each observation recorded
the composite spectrum of a one-degree diameter patch on the sky.
The spectral resolution of 12 km\thinspace s$^{-1}$~made it possible to
separate cleanly the interstellar emission from the terrestrial
emission in every spectrum and to measure the thermal and non-thermal
motions of the interstellar gas. Details about the WHAM-NSS and
downloadable versions of the survey can be found at {\tt
  http://www.astro.wisc.edu/wham/}. 

These survey maps of interstellar {H$\alpha$}~emission provide
the first global view of the distribution and motions of wide spread
ionized hydrogen within the Milky Way (see Figure 1).
The small bright knots of {H$\alpha$}~are classical emission nebulae
in the vicinities of hot O and B stars located mostly near the
Galactic midplane. Between these bright knots and filling most of the
sky is fainter {H$\alpha$}~emission from the Warm Ionized Medium
(WIM), with a characteristic temperature T$_e
\approx~$10$^4$\thinspace K and
density n$_e \approx~0.1$~cm$^{-3}$. 

Past studies have shown the WIM to be a significant
component of the interstellar medium, especially in the halos of disk
galaxies. In the Milky Way, the WIM has a mass surface density about
one-third that of neutral atomic hydrogen, a power requirement equal
to the kinetic energy injected into the Galaxy by supernovae, and a
characteristic scale height above the midplane of 1000 pc,
approximately five times larger than that of the neutral
hydrogen. This survey of the WIM shows rich structure both on the sky
and in velocity, allowing an exploration of the origin of the
ionization and heating of this gas and its relationship to the other
components of the interstellar medium. 

\section{High Latitude {H$\alpha$}~Emitting {H {\sc I}}~Clouds}

The intermediate velocity channel maps ($ -75~{\rm km\thinspace
  s^{-1}} \le v_{LSR} \le -50~{\rm km\thinspace s^{-1}}$) from the
survey reveal significant amount of high-latitude
{H$\alpha$}~emission.
Much of this emission is located near HI gas
previously identified and classified as Intermediate Velocity Clouds
(IVCs; see the recent compilation by Wakker 2001). One particularly
striking example is the IVC Complex K, reported by Haffner, Reynolds,
\& Tufte (2001).  

They find that general spatial correlation between
N$_{\rm{{H{\sc I}}}}$ and {H$\alpha$}~contours, and an excellent
kinematic correlation between {H$\alpha$}~and {H {\sc I}}~spectra
toward Complex K,  demonstrating that they probe the same structure. 
However, the spatial intensity distribution of the lines does not
appear to be related, i.e. peaks in {H {\sc I}}~emission do not
necessarily correspond to peaks in {H {\sc II}}~emission.  
This is consistent with photoionization by an external flux of
radiation. In this case, the {H$\alpha$}~intensity is determined by
the incident flux and is  unrelated to the {H~{\sc I}}~column density
of the cloud whose outer surface is being ionized. 

WHAM has also recently observed very faint {H$\alpha$}~emission
associated with the northern tip of the Magellanic Stream.  In all
five lines of sight observed, we find {H$\alpha$}~emission at  the
same velocity as the {H {\sc I}}~gas with intensities ranging between
0.05 - 0.1 R (EM $\approx 0.1 - 0.2~\rm{cm}^{-6}~\rm{pc}$ @ 10$^4$
K). Such observations are critical for our understanding of high and
intermediate velocity clouds, including the ability to constrain
models of the escape of Lyman continuum photons from the Galactic
disk (Bland-Hawthorn \& Maloney 1999). 

\begin{figure}
  \plotone{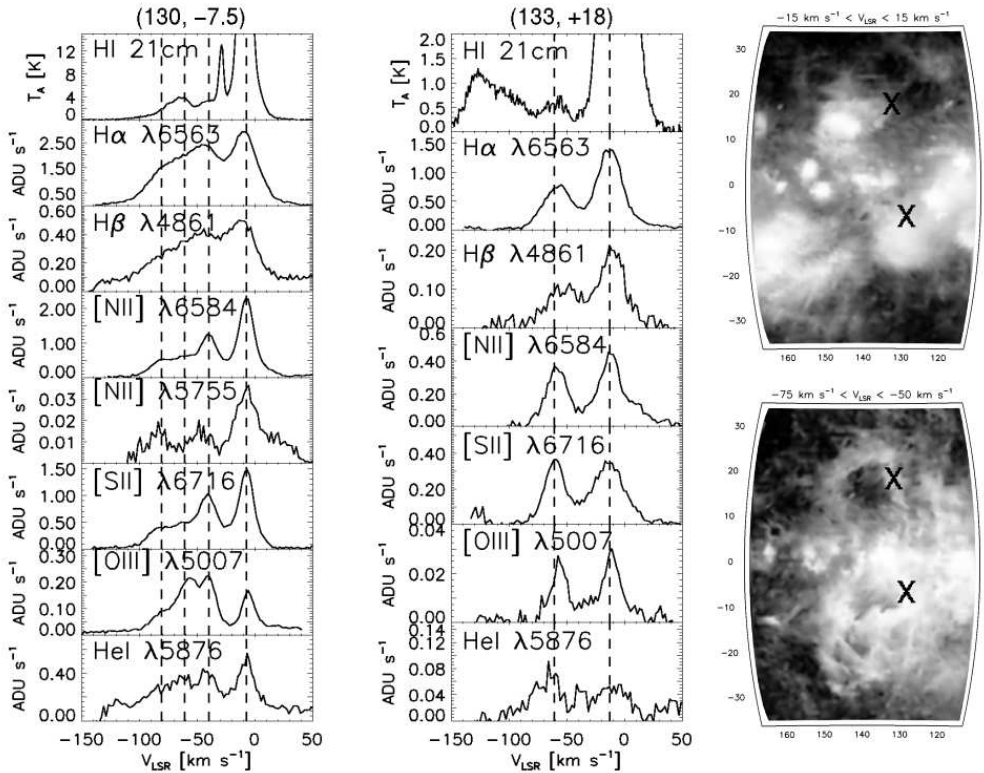}
  \caption{Multiwavelength spectra (left) and {H$\alpha$}~survey images
    (right) from WHAM. The
    spectra toward the two directions labeled by their
    titles are marked by an 'X' in the images on the right.
    Note the variations in the line ratios
    among the velocity components outlined by the dashed vertical
    lines in the spectra.
    The images on the right are two different velocity channel
    maps of the same region of the Galaxy, sampling local gas (top)
    and more distant Perseus arm gas (bottom).}
\end{figure}

\vspace*{-.1in}

\section{Probing the Heterogeneous Nature of the WIM}

The WHAM Northern Sky Survey has revealed the presence of remarkable
large-scale, {H$\alpha$}-emitting structures, including loops, filaments,
bubbles, and bright ``point sources'' throughout the WIM superposed on a
more diffuse {H$\alpha$}~background (e.g. Figure 1).
With the completion of the survey, WHAM can now be used to study the
WIM through observations of  several other optical emission 
lines, probing the physical properties of the gas, and with higher
angular resolution imaging, exploring its smaller scale structure. 

Figure 1 shows the spectra of sight lines towards $l=130\deg,
b=-7.5\deg$ and $l=133\deg, b=+18\deg$, sampling filamentary and
diffuse gas. The data are given in units of arbitrary flux versus LSR
velocity. 
Figure 1 also shows two velocity channel WHAM-NSS maps towards these
directions, chosen to isolate the local gas near the LSR and the more
distant gas in the Perseus arm near -60~km\thinspace s$^{-1}$. 
The 0~km\thinspace s$^{-1}$~frame is dominated by diffuse ionized gas
on which is superposed large, low density O and B star {H {\sc
    II}}~regions. The large ``bowtie'' in the Persues arm frame is
believed to be a superbubble blowout associated with the Cas OB6
association and the {H {\sc I}}~Normandeau ``chimney'' (Reynolds,
Sterling \& Haffner 2001; Normandeau, Taylor, \& Dewdney 1996). 

The spectra reveal multiple emission line components in each
direction, interpreted as local gas, Perseus spiral arm
gas a few hundred parsecs above the midplane at a distance of 2-3 kpc,
and higher velocity gas at even greater distances. We find significant
variations in the ratio 
of these emission lines among the different radial velocity components
along a single line of sight, as well as between the two lines of
sight. Preliminary analysis reveals interesting correlations and
anti-correlations among the line ratios and raises new questions about
the heterogeneous nature of the WIM.
Analysis of the strength of these emission lines and their ratios
reveal important clues about the temperature and ionization state of
the WIM, and indirectly reveal information about the ionizing
spectrum, extinction, density, and heating of the gas (Haffner,
Reynolds, \& Tufte 1999; Reynolds, Haffner, \& Tufte 1999).
We also find ionized gas associated with each of the warm
{H {\sc I}}~components toward (130,-7.5), but not with the cold
{H~{\sc I}}~feature near -30~km\thinspace s$^{-1}$. These observations  
can be used to explore how the physical conditions in the WIM change
with morphology, and how they compare with classical {H {\sc II}}~regions. 

Another new tool available with WHAM is higher angular resolution
imaging. With the insertion of a set of optics, WHAM can obtain very
narrow band, $3^{\prime}$ resolution images of the sky within its
1$\deg$ beam. An 
adjustable iris mechanism in the imaging optics can set the width of
the narrow spectral window of the image between 15~km\thinspace
s$^{-1}$~to 200~km\thinspace s$^{-1}$.
The combination of the multiwavelength observations of various
diagnostic emission lines with the higher angular resolution imaging
will allow us to learn more about the elusive nature of this important
phase of the interstellar medium. The WHAM Survey has already revealed
the rich spatial and kinematic structure of the warm ionized
medium. These multiwavelength and higher spatial resolution
observations will hopefully shed additional light on fundamental questions
pertaining to the nature of the WIM and no doubt will raise even more
questions.  This work was supported by National Science Foundation
Grant AST96-19424.

\end{document}